\documentclass{SCGE}
\usepackage{float}
\usepackage{graphicx}
\begin{document}

\begin{picture}(0, 0){\rm
\put(0, -20){\makebox[160truemm][l]{\bf {\sanhao\raisebox{2pt}{.}}
Article  {\sanhao\raisebox{1.5pt}{.}}}}}
\put(0, -34){\jiuwuhao {\textcolor[rgb]{0.5, 0.5, 0.5}{\sf 
}}}
\end{picture}

\def\bm{\boldsymbol}

\def\dl{\displaystyle}
\def\du{\end{document}}
\def\d{{\rm d}}
\def\e{{\rm e}}
\def\i{{\rm i}}

\Year{2015} %
\Month{January} %
\Vol{58} 
\No{3} 
\BeginPage{1} 
\EndPage{7} 
\AuthorMark{{\rm He J J },  et al.}  
\AuthorMarkCite{{\rm HE JianJun},  et al.} 
\DOI{} 

\title{A proposed direct measurement of cross section at Gamow window for key reaction $^{19}$F($p$,$\alpha$)$^{16}$O in Asymptotic Giant Branch stars with a
planned accelerator in CJPL}

\author[1]{HE Jianjun}{Corresponding author (email: jianjunhe@impcas.ac.cn)}
\author[1,2]{XU Shiwei}{}
\author[1,2]{MA Shaobo}{}
\author[1]{HU Jun}{}
\author[1]{ZHANG Liyong}{}
\author[3]{FU Changbo}{}
\author[1]{ZHANG Ningtao}{}
\author[4]{LIAN Gang}{}
\author[4]{SU Jun}{}
\author[4]{LI Yunju}{}
\author[4]{YAN Shengquan}{}
\author[4]{SHEN Yangping}{}
\author[1]{HOU Suqing}{}
\author[1,2]{JIA Baolu}{}
\author[3]{ZHANG Tao}{}
\author[3]{ZHANG Xiaopeng}{}
\author[4]{GUO Bing}{}
\author[1,5]{KUBONO Shigeru}{}
\author[4]{LIU Weiping}{}


\address[{\rm1}]{Institute of Modern Physics, Chinese Academy of Sciences, Lanzhou 730000, China;}
\address[{\rm2}]{University of Chinese Academy of Sciences, Beijing 100049, China;}
\address[{\rm3}]{Department of Physics, Shanghai Jiao Tong University, Shanghai 200240, People¡¯s Republic of China;}
\address[{\rm4}]{China Institute of Atomic Energy, P. O. Box 275(10), Beijing 102413, China;}
\address[{\rm5}]{RIKEN Nishina Center, 2-1 Hirosawa, Wako, Saitama 351-0198, Japan}

\maketitle \vspace{-3.5mm}{\footnotesize\begin{center} Received Month date, Year; accepted Month date, Year
\end{center}}\vspace*{-5mm}

\begin{center}
\rule{16.5cm}{0.4pt}
\parbox{16.5cm}
{\begin{abstract}
In 2014, the National Natural Science Foundation of China (NSFC) approved the Jinping Underground Nuclear Astrophysics laboratory (JUNA) project, which aims at
direct cross-section measurements of four key stellar nuclear reactions right down to the Gamow windows. In order to solve the observed fluorine overabundances
in Asymptotic Giant Branch (AGB) stars, measuring the key $^{19}$F($p$,$\alpha$)$^{16}$O reaction at effective burning energies (i.e., at Gamow window) is
established as one of the scientific research sub-projects. The present paper describes this sub-project in details, including motivation, status, experimental
setup, yield and background estimation, aboveground test, as well as other relevant reactions.
\end{abstract}}
\end{center}\vspace*{-0.6cm}

\begin{center}
\parbox{16.5cm}
{\bf\jiuhao  Jinping Underground Nuclear Astrophysics laboratory (JUNA), direct measurement, Gamow window, cross section, AGB star, nucleosynthesis}
\end{center}

\begin{center}
{}
\Cit{~~~He J J,  et al. {A proposed direct measurement of cross section at Gamow window for key reaction $^{19}$F($p$,$\alpha$)$^{16}$O in Asymptotic Giant Branch stars with a
planned accelerator in CJPL, Sci China-Phys Mech Astron, 2015, 58: 1--7, doi: }}
\end{center}

\textwidth=178truemm \textheight=236truemm

\wuhao\vspace*{1.5mm}

\begin{multicols}{2}

\renewcommand{\baselinestretch}{1.08} \baselineskip 12.2pt\parindent=10.8pt

\renewcommand{\thefootnote}

\section{Introduction}\label{sec:intro}
Nuclear processes play an extremely important role in the evolution of our Universe after the Big Bang~\cite{smith,mei}. Thereinto, nuclear reactions not only
provide the energy for stars to resist the gravitation, but also power the astrophysical explosion, such as x-ray bursts, novae and supernovae. Astrophysical
models that address the quiescent stellar evolutions and explosive astrophysical events require a huge amount of nuclear physics information as inputs.
Thermonuclear reaction cross section (or reaction rate) is one of the crucial quantities from nuclear physics aspect for modeling stellar phenomena. For
hydrostatic stable buring in stars, nuclear reactions occur at very low stellar energies. At the effective Gamow window, the extremely small cross sections
result in quite small signal-to-background ratio, which makes impossible the direct measurement in the laboratory at the Earth's surface. Covered by about
7000-mwe-thick marbles, China Jinping underground Laboratory (CJPL)~\cite{nor09,fed10,nor14}, the deepest underground laboratory in the world, can greatly
reduce the muon and neutron fluxes by 6 and 4 orders of magnitudes with respect to those at the Earth's surface. With such unique super-low-background and
salient features, the Jinping Underground Nuclear Astrophysics laboratory (JUNA) project was approved by the National Natural Science Foundation of China (NSFC)
in 2014 and will be financially supported in period of 2015--2019. The JUNA project$^*$\footnote{$^*$ Refer to five proposals submitted to NSFC in 2014, 
W.P. Liu et al., The $^{12}$C($\alpha$,$\gamma$)$^{16}$O reaction; X.D. Tang et al., The $^{13}$C($\alpha$,$n$)$^{16}$O reaction; Z. Li et al., The 
$^{25}$Mg($p$,$\gamma$)$^{26}$Al reaction; J.J. He et al., The $^{19}$F($p$,$\alpha_\gamma$)$^{16}$O reaction; L. Gang et al., The JUNA common platform. 
Please visit \textit{http://www.juna.ac.cn/pub/proposals} for details.} aims at direct measurement of ($\alpha$,$\gamma$) and ($\alpha$,$n$) reactions in hydrostatic helium burning, 
as well as ($p$,$\gamma$) ($p$,$\alpha$) reactions in hydrostatic hydrogen burning. In the first stage, four key reactions, i.e., 
$^{12}$C($\alpha$,$\gamma$)$^{16}$O, $^{13}$C($\alpha$,$n$)$^{17}$O, $^{25}$Mg($p$,$\gamma$)$^{26}$Al and $^{19}$F($p$,$\alpha$)$^{16}$O, will be directly 
measured at individual Gamow window~\cite{rol88}.

This paper describes details of the sub-project for $^{19}$F($p$,$\alpha$)$^{16}$O reaction study. The proposed experiment aims at direct cross section
measurement of this key stellar reaction right down to the Gamow window ($E_{c.m.}$=70--350 keV in the center-of-mass frame), with sufficient accuracy required
by the stellar model calculations. The direct experimental data will help people to expound the element abundances, especially the fluorine overabundances
observed in Asymptotic Giant Branch (AGB) stars, energy generation, as well as heavy-element nuclosynthesis scenario, with the astrophysical model on the firm
ground.

\section{Scientific motivation}\label{sec:motiv}
Fluorine is one of the most important elements for nuclear astrophysics. As the unique stable fluorine isotope, the $^{19}$F abundance is very sensitive to the
physical conditions within stars. Therefore, it is often used to probe the nucleosynthesis scenarios~\cite{luc11} of violent controversy. Most likely, fluorine
can be produced: 1) during core collapse of Type II supernovae~\cite{woo88}, 2) in Wolf-Rayet stars~\cite{mey00}, and 3) in the convective zone triggered by a
thermal pulse in Asymptotic Giant Branch (AGB) stars~\cite{cri09}. Recently, fluorine overabundances by factors of 800--8000~\cite{pan08} have been observed in
R-Coronae-Borealis stars, providing evidence for the fluorine synthesis in such hydrogen-deficient supergiants. However, a detailed description of fluorine
nucleosynthesis is still missing in despite of its crucial importance.

The major contributors to the Galactic fluorine~\cite{jor92} are the AGB stars. The observed fluorine overabundances cannot be explained with standard AGB models,
and additional mixing is still required~\cite{lug04}. For example, deep mixing phenomena in AGB stars can alter the stellar outer-layer isotopic composition
due to proton capture at low temperatures ($T_9$$\leq$0.04), and affect the transported material~\cite{nol03,ser10,bus10}. In this environment (corresponding to
Gamow window around $E_{c.m.}$=27--94 keV), the main fluorine destruction reaction of $^{19}$F($p$,$\alpha$)$^{16}$O
possibly modifies the fluorine surface abundances~\cite{luc11,abi11}. As for the hydrogen-deficient post-AGB stars, hydrogen admixture plays a key role to reverse
the effect of excessive He burning and yields elemental abundances in better agreement with observations~\cite{cla07}. Here, the $^{19}$F($p$,$\alpha$)$^{16}$O
reaction might bear a great importance as it would remove both protons and fluorine nuclei from the nucleosynthesis scenario. Therefore, the
$^{19}$F($p$,$\alpha$)$^{16}$O cross section should be well determined at $E_{c.m.}$$\sim$50--300 keV for accurate modeling because of the temperature at the base
of the accreted material approaching $T_9$$\sim$0.2~\cite{pan08}.

\begin{figure}[H]
\centering
\includegraphics[width=7.5cm]{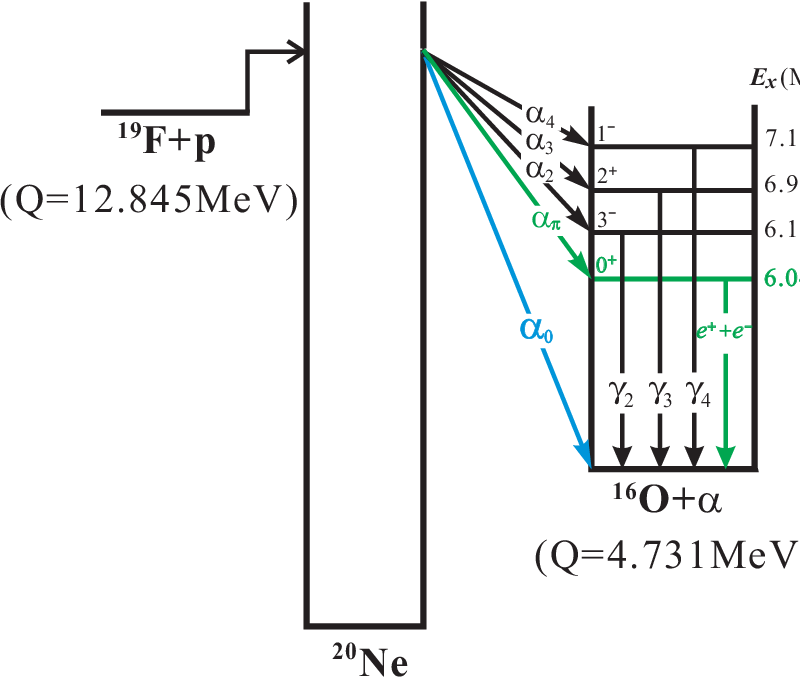}
\caption{Reaction scheme of $^{19}$F($p$,$\alpha$)$^{16}$O.}
\label{fig1}
\end{figure}

\section{Status and Goal}
The reaction scheme for $^{19}$F($p$,$\alpha$)$^{16}$O is shown in Fig.~\ref{fig1}. It shows that this reaction occurs through three different types of channels:
($p$,$\alpha_0$), ($p$,$\alpha_{\pi}$) and ($p$,$\alpha_{\gamma}$). Here, the combination of ($p$,$\alpha_2$), ($p$,$\alpha_3$) and ($p$,$\alpha_4$) with the
subsequent transitions of $\gamma_2$, $\gamma_3$ and $\gamma_4$, is referred to as the ($p$,$\alpha_{\gamma}$) channel. The research status for these reaction
channels are described in the following three subsections, and the final goal for the proposed experiment is summarized in the last subsection.

\subsection{($p$,$\alpha_0$) channel}
\begin{figure}[H]
\centering
\includegraphics[width=7cm]{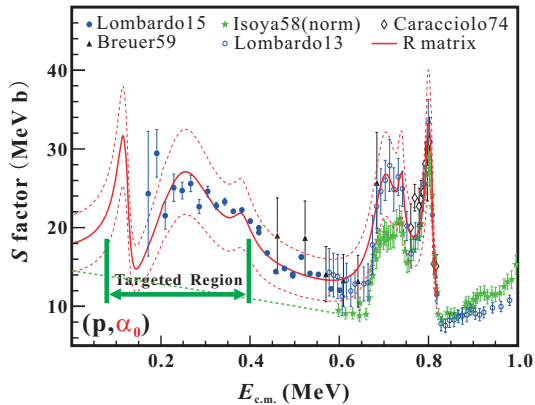}
\vspace{-3mm}
\caption{Available experimental data and $R$-matrix calculations for the ($p$,$\alpha_0$) channel~\cite{iso58,bre59,car74,lom13,lom15,ang99}. The green arrows
indicate the energy regions targeted for the JUNA experiment. (The figure is credited to I. Lombardo et al.)}
\label{fig2}
\end{figure}

The recommended $^{19}$F($p$,$\alpha_0$)$^{16}$O astrophysical $S(E)$-factor was determined from several works
~\cite{cla57,bre59,war63,car74,cuz80,iso56,iso58,isoy58,mor66} in the NACRE compilation~\cite{ang99}, with the lowest energy direct data at
$E_{c.m.}$=461 keV~\cite{bre59}. The Gamow window is only partially covered by the unpublished data of Lorentz-Wirzba~\cite{lor78}, which were
utilized~\cite{her91,yam93} later to evaluate the astrophysical factor in the zero and finite-range Distorted Wave Born Approximation (DWBA) approaches. These
data support a strong suppression of compound $^{20}$Ne decay to the ground state of $^{16}$O at $E_{cm}$$\sim$0.14--0.6 MeV. However, these results were not
included in the NACRE compilation as possible systematic errors affecting the absolute normalization might lead to an underestimate of $S(E)$ by a factor of
two~\cite{ang99}. The astrophysical factor was then extrapolated to low energies assuming a dominant contribution of the non-resonant part~\cite{ang99}. This
conclusion disagrees with the older measurement~\cite{bre59}, where the existence of two resonances with $J^{\pi}$=1$^-$ and 0$^+$ had been reported at
$E_{cm}$$\sim$0.4 MeV. Actually, additional resonances might be populated in $^{20}$Ne~\cite{til98}. A recent experiment~\cite{cog11} measured the
$^{19}$F($p$,$\alpha_0$)$^{16}$O astrophysical $S(E)$-factor by indirect means of the Trojan Horse method, and found that the largest rate difference, about 70\%,
occurs at temperatures relevant for post-AGB stars ($\sim$0.1 GK), exceeding the upper limit set by the previous uncertainties~\cite{ang99}. Such difference is
clearly due to the presence of the 113 keV resonance ($E_x$=12.957 MeV, 2$^+$). However, the energy resolution was not enough for achieving a good separation
between resonances, thus preventing an accurate estimate of the their total widths as well as the reaction rate. Most recently, Lombardo et al. reported new
direct measurement data~\cite{lom13,lom15} on the $^{19}$F($p$,$\alpha_0$)$^{16}$O reaction at the energy region of $E_{c.m.}$=0.18--1 MeV, and found the deduced
astrophysical $S$-factor $\approx$1.5--2 times larger than currently recommended one. However, their uncertainties are still too large below 0.2 MeV, which need
to be well constrained. The available experimental and theoretical data for this reaction channel are shown in Fig.~\ref{fig2}.

\subsection{($p$,$\alpha_\gamma$) channel}
\begin{figure}[H]
\centering
\includegraphics[width=8cm]{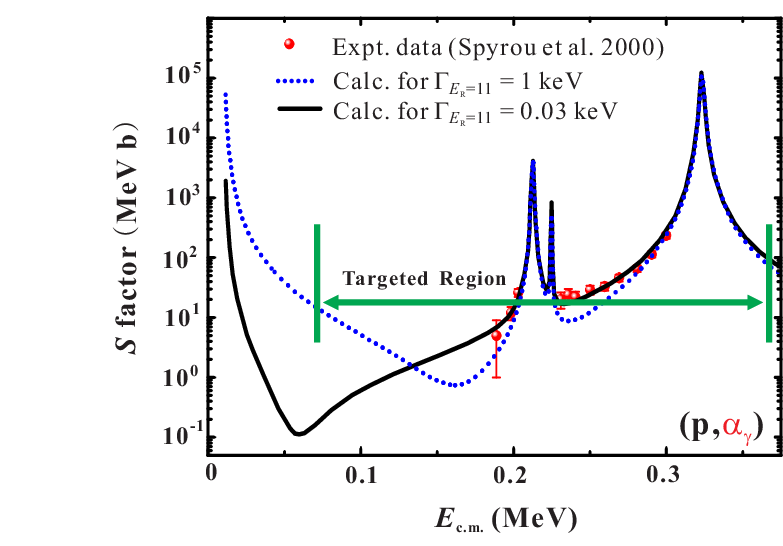}
\vspace{-3mm}
\caption{Available experimental data and theoretical calculations for the ($p$,$\alpha_\gamma$) channel~\cite{ang99,spy00}. The green arrows indicate the energy
regions targeted for the JUNA experiment. (The figure is credited to K. Spyrou et al.)}
\label{fig3}
\end{figure}

The recommended $^{19}$F($p$,$\alpha_\gamma$)$^{16}$O astrophysical $S(E)$-factor in the NACRE compilation~\cite{ang99} had been derived from the earlier
works~\cite{bon48,cha50,ask65,bec82,gra84,cro91,zah95,spy97} down to $E_{c.m.}$=957 keV. Later on, Spyrou et al.~\cite{spy00} measured this channel down to
$\sim$189 keV with a 4$\pi$ NaI summing spectrometer. In this work, the strengths of all resonances at $E_p$=200--800 keV, including a new one at $E_R$=237 keV,
have been extracted. Furthermore, the width of the important 1$^+$ resonance at $E_R$=11 keV was estimated, which affects the $S$-factor dramatically within the
Gamow region owing to the inference effects with the strong 1$^+$ resonance at $E_R$=340 keV. Therefore, this width needs to be determined experimentally.
The available experimental and theoretical data for this reaction channel are shown in Fig.~\ref{fig3}.

\subsection{Roles of different channels}
\begin{figure}[H]
\centering
\includegraphics[width=8cm]{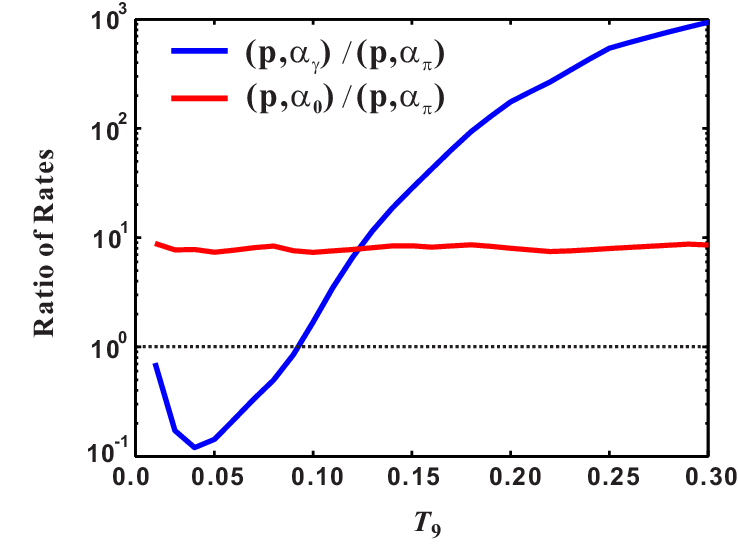}
\caption{The role of three reaction channels contributing to the total reaction rate of $^{19}$F($p$,$\alpha$)$^{16}$O.}
\label{fig4}
\end{figure}

Based on the work of Spyrou et al.~\cite{spy00} and the NACRE compilation~\cite{ang99}, the roles of three reaction channels are compared as shown in
Fig.~\ref{fig4}. It shows that the ($p$,$\alpha_0$) channel dominates the total rate below $\sim$0.1 GK; the ($p$,$\alpha_\gamma$) channel dominates above
$\sim$0.15 GK; both two channels dominates over 0.1--0.15 GK. Here, contribution from the ($p$,$\alpha_\gamma$) channel is negligible based on our current
knowledge about this reaction. Therefore, we will focus on the measurements of ($p$,$\alpha_0$) and ($p$,$\alpha_\gamma$) channels in the proposed experiment.

\subsection{Final goal}
In a summary, there are currently no direct experimental data below $E_{c.m.}$=180 keV for the ($p$,$\alpha_0$) channel (note: the uncertainty of the datum
point at 180 keV is quite large in Ref.~\cite{lom15}), and below $E_{c.m.}$=189 keV for the ($p$,$\alpha_\gamma$) channel (note: the uncertainties of the data
points below 200 keV are quite large in Ref.~\cite{spy00}). In the proposed experiment, we will target on measuring the cross sections of these two reaction
channels at the targeted energy regions indicated in Figs.~\ref{fig2}~\&~\ref{fig3} in JUNA. As a final goal, we may obtain the reliable direct experimental
data with a precision about 10\% at lower energies or even better at higher energies (see Tabs.~\ref{alpha}~\&~\ref{gamma yield}), and implement them into the
nucleosynthesis model to achieve a better understanding of the fluorine overabundances in AGB stars.

\section{Experimental setup}
In the proposed $^{19}$F($p$,$\alpha$)$^{16}$O experiment, two reaction channels of ($p$,$\alpha$) and ($\alpha$,$p$) will be measured separately. The details for
these two measurements will be described as below.

\begin{figure}[H]
\centering
\includegraphics[width=7cm]{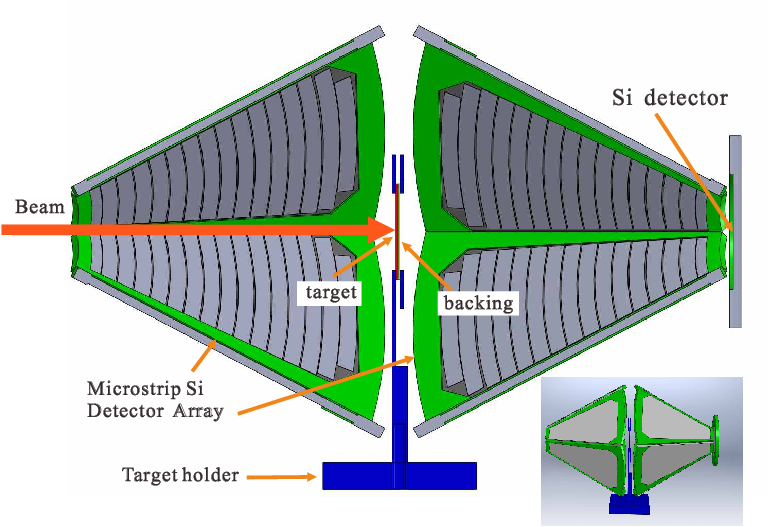}
\caption{Conceptual silicon detector array designed for measuring the charged particles.}
\label{fig5}
\end{figure}

A `lamp'-type Micron silicon array will be constructed for the charged-particle measurement, which can cover about 4$\pi$ solid angle. This
universal detection array will set the base for other ($p$,$\alpha$) and ($\alpha$,$p$) reaction studies at JUNA. A conceptual design is shown in Fig.~\ref{fig5}.
It can not only measure the total ($p$,$\alpha_0$) cross section but also the angular distribution. The angular distribution measured is much useful for
determining the nuclear structure of the 1$^+$ resonance at $E_R$=11 keV as discussed above. In this experiment, a very thin about 4~$\mu$g/cm$^2$ CaF$_2$ target
will be utilized, which is evaporated on a thin backings. Thanks to the very high $Q$ value (about 8.11 MeV) for this reaction, the average energy for the
emitted $\alpha$ particles is about 6.7 MeV. These relatively high-energy $\alpha$s can penetrate the backings and can be detected easily at the forward angle.
The detectors at the forward angle do not face the Rutherford-scattered strong proton beam which is stopped in the backings. However, those detectors at the
backward angle should be shielded by a thin foil, e.g., a mylar foil, to stop the scattered protons. The target backing will be connected to a cooling device to
release the heat during the experiment.

\begin{figure}[H]
\centering
\includegraphics[width=8cm]{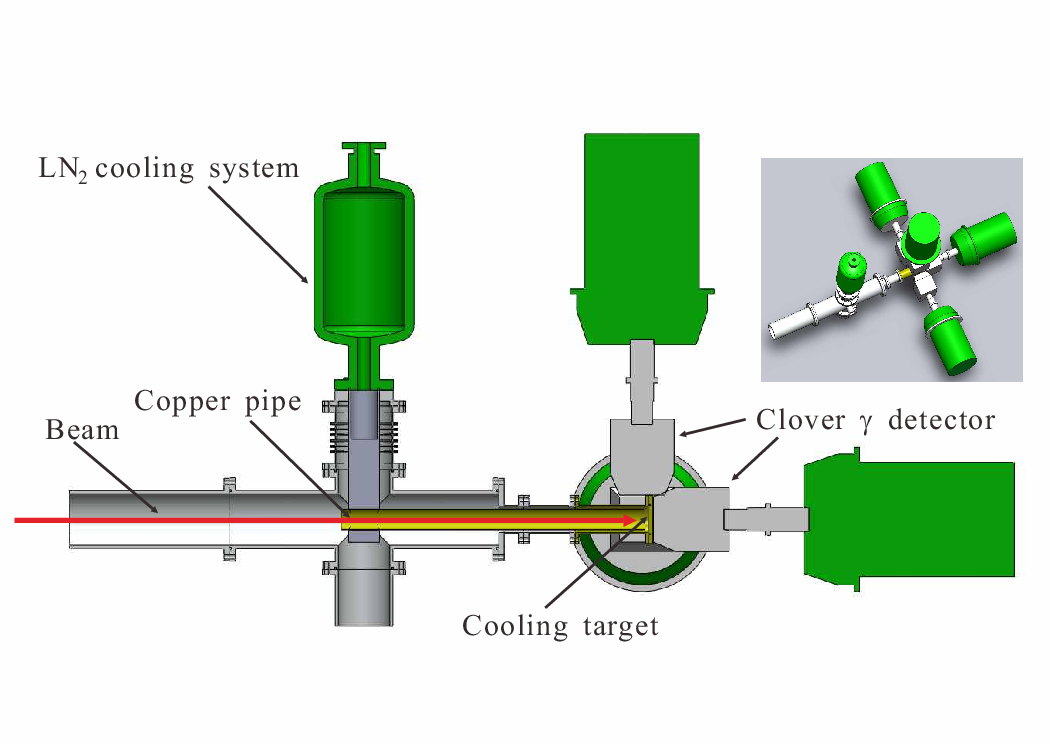}
\caption{Conceptual Clover detector array designed for measuring the $\gamma$ rays.}
\label{fig6}
\end{figure}

As for the $^{19}$F($p$,$\alpha_\gamma$)$^{16}$O channel, the energies of emitted $\gamma$ rays are about 6--7 MeV, where the background mainly originates from
the cosmic rays. Covered with about 2400 km marbles, such background can be greatly reduce, which makes the low-energy measurements feasible. In this project,
two gamma detection arrays will be constructed for the $\gamma$-ray measurements. One is the High-Purity Germanium (HPGe) Clover array whose absolute detection
efficiency is about 1\%~\cite{Chen} for the $\gamma$ rays of interest, but with excellent energy resolution; another is the 4$\pi$ BGO array whose absolute 
efficiency is estimated to be about 75\%, but with relatively worse resolution. Now, this BGO array is under construction. Here, the Clover array will be 
utilized in the $E_{c.m.}$$>$140 keV energy region, while the BGO array will be used below this energy region. With the excellent resolution of the Clover
detector, the possible contaminations can be resolved and identified clearly, which makes the BGO $\gamma$-ray summing reliable at lower-energy region.
A conceptual design for the Clover array is shown in Fig.~\ref{fig6}.

\section{Yield estimation}
\noindent \emph{Yield estimated for the ($p$,$\alpha$$_0$) channel:}

Based on the theoretically calculated cross section in Ref.~\cite{lom15}, the alpha yields from the ($p$,$\alpha_0$) channel have been estimated in
Table~\ref{alpha}. In the calculations, we propose to use a 4~$\mu$g/cm$^2$ thickness CaF$_{2}$ target in which the energy loss of a 70-keV proton is
only about 1.6 keV. The detection efficiency of the `lamp'-type silicon array is estimated to be about 80\%. It can be seen from Table~\ref{alpha} that
measurements below 90~keV energy region become increasingly difficult because of the much smaller cross section. As a conservative estimation, the highest
current utilized in the present calculation is 100 $\mu$A. The experiment becomes easier with the higher current, but the target must be cooled accordingly.
Based on the experience of LUNA experiments, the directly water cooled target can fully endure the supposed 100 $\mu$A current.

\end{multicols}

\begin{table}
\caption{\label{alpha} $\alpha$ yields estimated for the ($p$,$\alpha_0$) channel. The $S$-factor data are estimated from the (red) solid
line in Fig.~\ref{fig2}, please refer to Ref.~\cite{lom15}.}
\begin{center}\footnotesize \doublerulesep 0.2pt \tabcolsep 22pt
\begin{tabular}{cclcr}\toprule[0.65pt]
$E_{c.m.}$ (keV) & $S$-factor (MeV b)~\cite{lom15}  & Cross section (b) & Current ($\mu$A)  &  Counting rate  \\
\hline
70  & 21.0 & 1.70$\times$10$^{-12}$ (pb) & 100   & 47$\slash$week     \\
80	& 22.0 & 1.29$\times$10$^{-11}$	     & 100   & 360$\slash$week    \\
90	& 23.5 & 7.10$\times$10$^{-11}$	     & 100   & 280$\slash$day     \\
100	& 26.5 & 3.18$\times$10$^{-10}$	     & 100   & 1270$\slash$day    \\
130	& 20.0 & 5.40$\times$10$^{-9}$ (nb)  & 10	 & 2160$\slash$day    \\
160	& 16.0 & 3.77$\times$10$^{-8}$	     & 1	 & 1500$\slash$day    \\
180	& 18.5 & 1.34$\times$10$^{-7}$	     & 1	 & 220$\slash$hour    \\
200	& 22.0 & 4.10$\times$10$^{-7}$	     & 1	 & 680$\slash$hour    \\
250	& 27.0 & 3.12$\times$10$^{-6}$	     & 1	 & 5200$\slash$hour   \\
350	& 21.5 & 2.61$\times$10$^{-5}$	     & 0.1   & 4350$\slash$hour   \\
\bottomrule[0.65pt]
\end{tabular}
\end{center}
\end{table}

\begin{table}
\caption{\label{gamma yield} $\gamma$ yields estimated for the ($p$,$\alpha_\gamma$) channel. The $S$-factor data are estimated from the solid
line in Fig.~\ref{fig3}, please refer to Ref.~\cite{spy00}.}
\begin{center}\footnotesize \doublerulesep 0.2pt \tabcolsep 15pt
\begin{tabular}{cclcrc}\toprule[0.65pt]
$E_{c.m.}$ (keV) & $S$-factor (MeV b)~\cite{spy00} & Cross section (b)  & Current ($\mu$A)	&  Counting rate  & Detector ($\epsilon_\gamma$) \\
\hline
100	& 0.6 & 7.21$\times$10$^{-12}$ (pb)  & 100   & 27$\slash$day	 & BGO  (75\%) \\
110	& 0.8 & 3.13$\times$10$^{-11}$	     & 100   & 118$\slash$day	 & BGO  (75\%) \\
120	& 1.0 & 1.09$\times$10$^{-10}$	     & 100   & 410$\slash$day    & BGO  (75\%) \\
140	& 1.8 & 1.08$\times$10$^{-9}$ (nb)   & 100   & 54$\slash$day	 & HpGe (1\%)  \\
160	& 3.5 & 8.24$\times$10$^{-9}$  	     & 50	 & 206$\slash$day	 & HpGe (1\%)  \\
200	& 12  & 2.23$\times$10$^{-7}$	     & 20	 & 93$\slash$hour	 & HpGe (1\%)  \\
250	& 29  & 3.35$\times$10$^{-6}$	     & 5	 & 350$\slash$hour	 & HpGe (1\%)  \\
300	& 235 & 1.03$\times$10$^{-4}$	     & 1	 & 2140$\slash$hour	 & HpGe (1\%)  \\
\bottomrule[0.65pt]
\end{tabular}
\end{center}
\end{table}

\begin{multicols}{2}
\renewcommand{\baselinestretch}{1.08} \baselineskip 12.2pt\parindent=10.8pt

\noindent \emph{Yield estimated for the ($p$,$\alpha_\gamma$) channel:}
Based on the theoretically calculated cross section in Ref.~\cite{spy00}, the $\gamma$-ray yields from the ($p$,$\alpha_\gamma$) channel have been estimated
as listed in Table~\ref{gamma yield}. This reaction becomes dominant above 0.12 GK (see Fig.~\ref{fig4}), and the corresponding lower limit of the Gamow energy
is $\sim$100 keV. Similarly, the target thickness is also assumed to be 4~$\mu$g/cm$^2$ in the calculation. It can be seen from Table~\ref{gamma yield} that
even at the lowest 100 keV (cross section at the order of pb), 380 counts can be expected in two weeks, with a statistical error of $\sim$5\%.

It should be noted that the JUNA 400~kV accelerator can provide beam current in the order of mA. This strong beam will mainly apply to those those reactions with
extremely low cross section reactions (e.g., $^{12}$C($\alpha$,$\gamma$)$^{16}$O) and targets with very high melting point (such as the melting point is about
3500 $^{\circ}$C for the C target). As for the CaF$_{2}$ target, the melting point is only about 1400 $^{\circ}$C. Here, we only assumed a 100 $\mu$A beam
current. The practical highest current which the target can stand will be tested experimentally.

\begin{figure}[H]
\centering
\includegraphics[width=8cm]{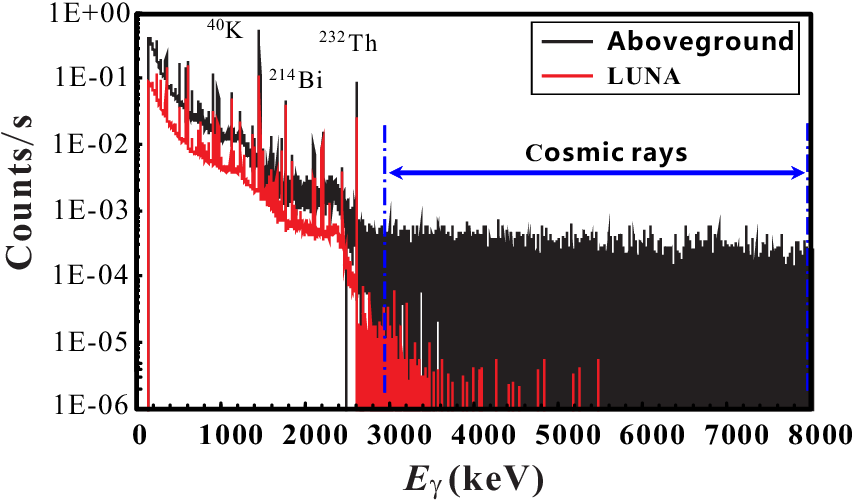}
\caption{Comparison of $\gamma$-ray background levels at LUNA and earth surface~\cite{Imbriani}.}
\label{fig7}
\end{figure}

\section{Background estimation}
\noindent \emph{Charged-particle background:}
In the deep underground laboratory, the environmental and material background become dominant in the charged-particle detection. All of the material used in the
experiment, such as target chambers, targets and the holders, detectors and even the radiation of the shielding material can disturb the charged-particle
detection~\cite{Iliadis}. For instance, the $\alpha$ activity of the stainless steel is one order of magnitude lower than that of the commercial aluminum material.
Moreover, cosmic-ray induced $\gamma$ rays, neutrons and charged particles can also affect the rare event detection. (1) In the $^{3}$He($^{3}$He,2$p$)$^{4}$He
experiment at LUNA, the cosmic-ray induced background in the silicon detectors is as less as 3.5$\times$10$^{-4}$ event/s, which is about 200 times lower
than that achieved aboveground~\cite{Greife}. The JUNA background is expected to be better than that of LUNA because of the depth. (2) The neutron background in
LUNA is about 3 orders of magnitude lower than that aboveground, and additionally the JUNA background is about 10 times~\cite{LUNA} lower than that of LUNA.
Therefore, the influence of neutron background on Si detectors at JUNA is about 4 orders of magnitude lower than that aboveground. (3) It was measured that the
radioactivities of the surrounding rocks in JUNA is much lower than the earth surface level (see Table~\ref{alpha_back}). These natural $\alpha$ radioactivities
(together with $\gamma$ radiation) will produce background in Si detectors. Furthermore, the Si detectors in the target chamber should be shielded by the
\emph{old lead} or oxygen-free copper with extremely low background. Moreover, the $\mu$ particles in cosmic ray can only deposit less than 0.5 MeV energy
in the detectors, while the targeted $\alpha$ particles have energy loss more than 6 MeV, and therefore, the extremely low charged-particle background in
JUNA makes the present measurement around 70 keV feasible.

\end{multicols}

\begin{table}
\caption{\label{alpha_back} Comparison of the radioactivity background for the surrounding rocks in JUNA, LUNA and earth surface. The data of LUNA are taken 
from Ref.~~\cite{Imbriani}.}
\begin{center}\footnotesize \doublerulesep 0.2pt \tabcolsep 22pt
\begin{tabular}{ccccc}\toprule[0.65pt]
Location	  & $^{40}$K  & $^{222}$Rn (Bq$\slash$m$^3$) & $^{226}$Ra (Bq$\slash$kg)  & $^{232}$Th (Bq$\slash$kg)    \\
\hline
JUNA	      & $<$~0.1~\cite{Kang}    & 10--20~\cite{Cao}	    & 1.8$\pm$0.2~\cite{Kang}	 & $<$~0.27~\cite{Kang}  \\
LUNA	      & 224       & 20--90	&  -	                     & 8.8      \\
Beijing ground level & 600~\cite{Cao}  & 200--400	            & 25~\cite{Cao}	             & 50~\cite{Cao}         \\
\bottomrule[0.65pt]
\end{tabular}
\end{center}
\end{table}

\begin{table}
\caption{\label{gbk} Comparison of the $\gamma$-ray background levels for laboratories of JUNA, LUNA and earth surface.}
\begin{center}\footnotesize \doublerulesep 0.2pt \tabcolsep 22pt
\begin{tabular}{ccrr}\toprule[0.65pt]
Laboratory	&    Depth (km)	&    Cosmic-ray flux (cm$^{-2}$s$^{-1}$)	&   Counting rate of 3--8 MeV $\gamma$ rays \\
\hline
JUNA	        & 2400	    & 2$\times$10$^{-10}$~\cite{Wu}	        & 2$\times$10$^{-6}$$\slash$s (estimated)       \\
LUNA	        & 1400	    & 3$\times$10$^{-8}$~\cite{Imbriani}	& 2$\times$10$^{-4}$$\slash$s~\cite{Imbriani}   \\
Earth surface	& $\sim$0	& 2$\times$10$^{-2}$~\cite{Heusser} 	& 0.5$\slash$s~\cite{Imbriani}                  \\
\bottomrule[0.65pt]
\end{tabular}
\end{center}
\end{table}

\begin{multicols}{2}
\renewcommand{\baselinestretch}{1.08} \baselineskip 12.2pt\parindent=10.8pt

\noindent \emph{$\gamma$-ray background:}
JUNA will provide an unique ultra-low background level in the world, which makes the rare-event detection possible. The 3--8 MeV $\gamma$-ray background in LUNA
is about 2$\times$10$^{-4}$ event/s~\cite{Imbriani}, $\sim$2000 times lower than that aboveground, which is mainly caused by the cosmic rays (see Fig.~\ref{fig7}).
Owing to the depth advantage, the cosmic-ray induced background at JUNA is expected to be about 100 times lower than that in LUNA~\cite{Wu} (see also
Table~\ref{gbk}). Thus, the 3--8 MeV $\gamma$-ray background at JUNA is estimated to be 2$\times$10$^{-6}$ event/s, i.e., 0.17 event/day. By taking the
beam-induced background into account, the $\gamma$ background is expected to be about 0.25 event/day. Based on the above estimation, the targeted
$\gamma$-ray yield at 100 keV is 27 events/day, far greater than the background level. In aboveground lab., the background level is 0.5 event/s, i.e.
4.3$\times$10$^{5}$ events/day, which indicates only the measurements above 200 keV can be performed~\cite{spy00}. Consequently, JUNA can provide us an
excellent condition to extend the $^{19}$F($p$,$\alpha_\gamma$)$^{16}$O cross section measurement down to the Gamow window. For comparison, the background levels
in LUNA, JUNA and earth-surface lab. are listed in Table~\ref{gbk}. Recently, the realistic $\gamma$-ray background has been measured by both a HPGe and a BGO
crystal at JUNA, and the data analysis is still in progress.

Comparing to the radioactivity background for the surrounding rocks in JUNA, the cosmic-ray induced $\gamma$ background can be entirely neglected because of the
extremely reduced muon fluxes at JUNA (see Table~\ref{gbk}). Therefore, a good shielding for the background originated from the surround rocks, concrete wall, as
well as accelerator is quite important for the underground experiments. In addition, the radioactive decay products from the radon (existing in the air and rocks)
can pollute the surface of the detector (in the order of about 1 event/month), and hence the $\gamma$-ray detectors should be assembled, disassembled and operated
in the radon free environment. Furthermore, the decaying radon and its daughters produce $\alpha$ and $\beta$ particles that produce again secondary $\gamma$
radiation by bremsstrahlung and nuclear reactions. A popular solution of this problem is to house the detector in a box with a small overpressure of flushing
nitrogen: i.e., by substituting normal air containing Radon with Nitrogen inside the box~\cite{LUNA}.



\section{Further studies}
In nuclear-physics input aspect, it is necessary to well study other three keys reactions in order to solve the observed fluorine overabundances in AGB stars,
except the $^{19}$F($p$,$\alpha$)$^{16}$O reaction proposed above. These are the $\alpha$-induced reactions of
$^{13}$C($\alpha$,$n$)$^{17}$O~\cite{ang99,guo12,cog12}, $^{14}$C($\alpha$,$\gamma$)$^{18}$O~\cite{joh09} and $^{19}$F($\alpha$,$p$)$^{22}$Ne~\cite{uga08}.
As one of the experimental goals in the JUNA project, the details of the first ($\alpha$,$n$) reaction is described elsewhere~\cite{tang}.
We will target on studying the remaining two important reactions following the proposed $^{19}$F($p$,$\alpha$)$^{16}$O experiment, with the $\gamma$-ray and
charged-particle detector arrays constructed. Certainly, such universal charged-particle array can play a very important role in JUNA, to study other key
reactions (emitting charged particles) of astrophysical importance.

\vspace*{2mm} \Acknowledgements{\bahao This work was financially supported by the National Natural Science Foundation of China (Nos. 11490562, 11490560).}

\end{multicols}
\end{document}